**Decoupling Intrinsic Molecular Efficacy from Platform Effects: An Interpretable Machine Learning Framework for Unbiased Perovskite Passivator Discovery**

*Jing Zhang, Ziyuan Li, Shan Gao, Zhen Zhu, Jing Wang*, and Xiangmei Duan**

School of Physical Science and Technology, Ningbo University, Ningbo, China
E-mail: wangjing4@nbu.edu.cn; duanxiangmei@nbu.edu.cn

Funding: This work is supported by the National Natural Science Foundation of China (Grant No. 12374061, No. 12404274, No. 12204256), Zhejiang Provincial Natural Science Foundation of China (Grant No. LQN25A040020, No. LD24F040001), the KC Wong Magna Foundation in Ningbo University. The computations were supported by high performance computing center at Ningbo University.

**Abstract**

Rational design of interface passivators for perovskite solar cells is hindered by the entanglement of intrinsic molecular efficacy with extrinsic platform-dependent performance—a confounding factor that obscures true chemical advances. Here, we present a generalizable, interpretable machine learning framework that decouples these effects via an asymptotic saturation model, enabling unbiased discovery of molecules with genuine intrinsic gains. Trained on a curated dataset of 240 experimental entries, our model identifies hydrogen bond acceptor strength and electrostatic potential difference as key descriptors. Guided by these insights, we screened >121 million PubChem compounds using a hierarchical strategy integrating diversity clustering and uncertainty quantification. Five dual-functional candidates (e.g., TDZ-S, TZC-F) are identified, exhibiting superior predicted efficacy (surpassing experimental benchmarks) and high confidence. First-principles calculations confirm strong chemisorption ($E_{ads} < -1.7$ eV), net electron donation, and optimized interfacial energetics. Crucially, our closed-loop "data–interpretation–screening–verification" pipeline establishes a transferable paradigm for rational materials design, extendable to other optoelectronic interfaces beyond perovskites.

# 1. Introduction

Hybrid perovskites have emerged as a foremost promising materials for next-generation photovoltaic technologies, propelled by their adjustable bandgaps, cost-effective fabrication processes, and outstanding optoelectronic characteristics.[1-4] Recent advancements have pushed the certified efficiencies of single-junction perovskite solar cells (PSCs) beyond 27 %,[5] nearing the performance levels of traditional silicon cells. However, the inherent softness of the perovskite lattice leads to the formation of structural defects at both surface and interfaces.[6] These defects typically serve as non-radiative recombination centers, causing increased voltage losses, reduced device efficiency, and accelerated degradation under operational conditions.[7] Therefore, managing these defects is essential for the commercial realization of PSC technology.[8-12]

Interface passivation has emerged as a pivotal strategy for mitigating these losses, thanks to its localized efficacy and process compatibility.[13-15] A diverse array of passivating agents, ranging from organic small molecules to polymers, have demonstrated the ability to neutralize deficient $Pb^{2+}$ ions or halide vacancies. This is achieved through Lewis acid-base interactions, hydrogen bonds, and electrostatic coupling facilitated by functional groups such as ammonium or carboxyl groups.[16-24] Despite these advancements, the development of efficient passivating agents remains heavily reliant on empirical methods. Yet conducting trial-and-error exploration within a vast chemical space is inefficient, and it difficult to disentangle complex, non-linear synergies among molecular attributes using traditional single-variable approaches.[25,26] Consequently, there is an urgent need for a rational design framework that bridges the gap between chemical structure and device performance. Notably, while single-functional passivators (e.g., targeting only Lewis acids) have shown promise, they often fall short in addressing the complex, multi-defect landscape of perovskite surfaces. Developing "dual-functional" molecules capable of simultaneously neutralizing Lewis acidic and basic defects through synergistic interactions represents a more robust, yet underexplored, avenue for maximizing passivation efficacy.[27-30]

Machine learning (ML) offers a potent tool to tackle these high-dimensional complexities and now serve as cutting-edge instrument for materials discovery.[31-33] In the realm of perovskite photovoltaics, ML has been successfully harnessed for a spectrum of applications, including composition optimization, stability forecasting, and absorber layer screening.[34-40] Nevertheless, a critical gap persists: Many existing ML studies lean heavily on generic molecular fingerprints that lack direct relevance to the specific defect chemistry of perovskites.[41-44] Moreover, the opaque nature of many models—functioning as "black

boxes"—obscures the internal decision logic, thereby limiting their utility for rational, insight-driven design.[45]

Herein, we establish a closed-loop workflow integrating interpretable ML with theoretical verification to accelerate the discovery of efficient passivators. A feature set tailored for the passivation process was constructed, which was used to train a high-performance random forest (RF) model (test $R^2$=0.914). Using SHapley Additive exPlanations (SHAP) analysis,[46] we decoded the model to reveal non-linear, context-dependent interactions among key molecular features and, crucially, to decouple the intrinsic efficacy of passivators from platform-dependent baseline effects. Guided by these insights, we screened the PubChem database (>121 million molecules) using a hierarchical strategy incorporating uncertainty quantification to identify five high-confidence candidates. First-principles calculations confirmed that these molecules possess strong adsorption, favorable band alignment, and net electron-donor characteristics. In summary, this work established a complete "data mining—model–interpretation–screening–validation" loop, providing a transferable paradigm for the rational discovery of complex interface materials.

## 2. Results and Discussion
### 2.1. Data-Driven Framework and Feature Engineering

We introduce a meticulously curated data-driven workflow that synergistically integrates data mining, machine learning, and theoretical verification to accelerate the discovery of efficient perovskite interface passivators. As depicted in **Figure 1**, the workflow comprises five consecutive steps forming a closed loop. (1) Data Collection: A high-quality dataset was curated from a systematic literature review spanning the past decade. To target high-efficiency photovoltaics, we compiled 240 entries involving 218 unique organic passivators (**Figure S1, Table S1**), strictly filtering for Pb-based devices with initial efficiencies >18 %. (2) Machine Learning Modeling: Custom features were extracted to train regression models (e.g., Random Forest) for robust structure-property mapping. (3) Interpretability Analysis: The optimal model was dissected using multi-level SHAP analysis, progressing from global importance to local attribution for mechanistic insight. (4) Virtual Screening: The model was applied to the PubChem database to identify candidates via hierarchical filtering and uncertainty quantification. (5) First-Principles Verification: Top candidates were validated using Density Functional Theory (DFT) to probe electrostatic and adsorption properties.

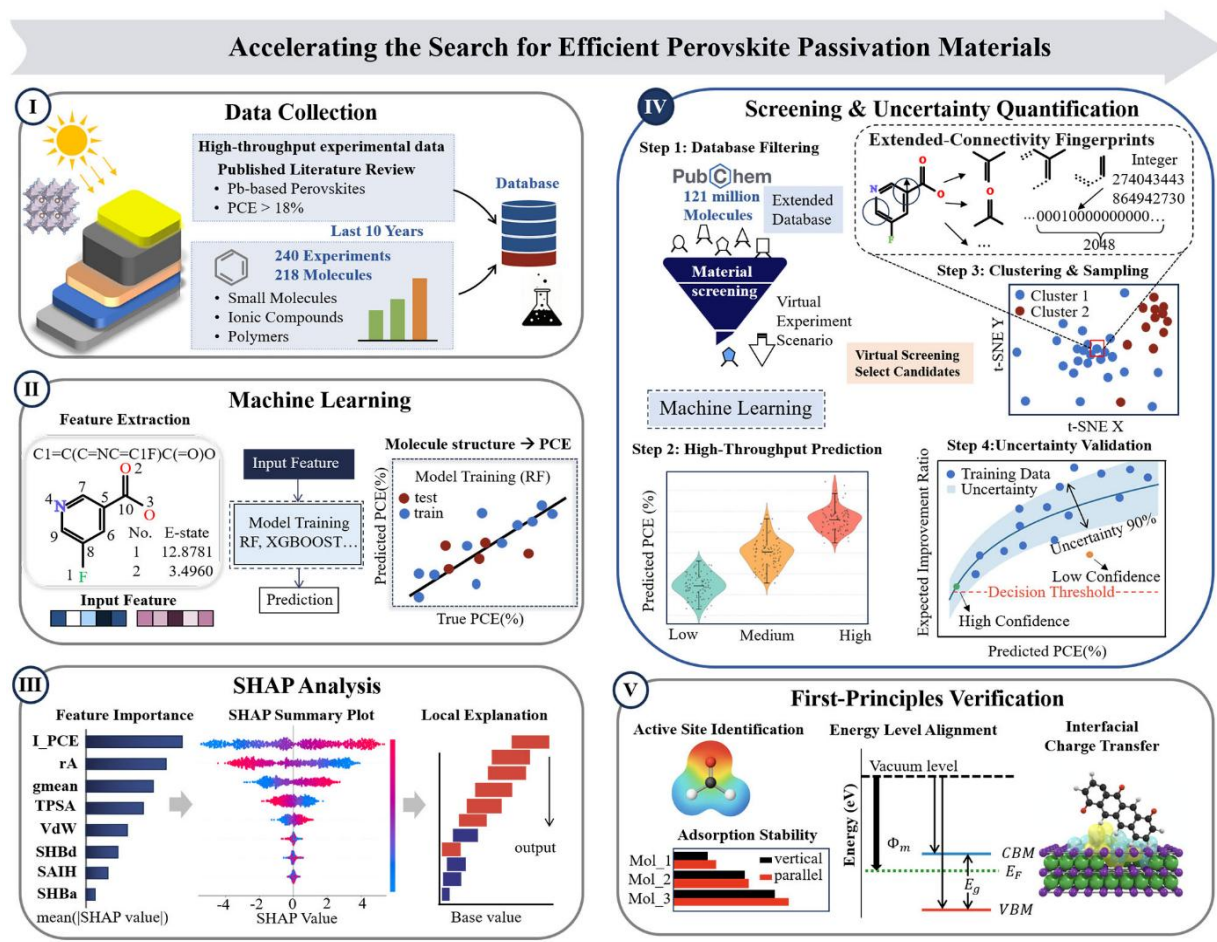

**Figure 1.** Schematic illustration of the closed-loop data-driven workflow developed in this study. The workflow comprises five subsequential steps: I) Data Collection and Dataset Construction; II) Machine Learning Modeling; III) SHAP-based Interpretability Analysis; IV) Screening and Uncertainty Quantification; and V) First-Principles Verification.

Preliminary statistical analysis reveals that both surface and interface passivation ubiquitously improve photovoltaic parameters, with the power conversion efficiency (PCE) displaying the most pronounced sensitivity to structural variations (**Figure S2**). Consequently, PCE was selected as the prediction target. To map molecular structures to performance, a 21-dimensional feature set was engineered to capture the physicochemical essence of passivation (**Table 1**). Firstly, initial device efficiency (I_PCE) was introduced as a baseline to decouple intrinsic film quality from passivation effects. This was complemented by the effective cation radius (rA) and precursor molar ratios (A:X, B:X), which critically influence crystallization dynamics and defect formation.[47] Secondly, considering charge interactions, dipole moments (u, uD) and Gasteiger charge descriptors (gmean, qpmax, qnmax) were incorporated to quantify electronic distribution.[48,49]

In addition, electro-topological state indices (E-state) were computed to assess the defect anchoring capabilities, specifically hydrogen bond donor (SHBd) and acceptor (SHBa) properties.[50] These features quantify chemical adsorption potential by aggregating functional group contributions.[51-53] To further address steric and interface compatibility, the rotatable bond fraction (RBR) was incorporated to reflect molecular flexibility, while the topological polar surface area (TPSA) and partition coefficient (LogP) were employed to capture polarity and hydrophobicity characteristics.[54-[55]56] Moreover, the van der Waals volume (vdW) and complexity index (CPLX) were introduced to comprehensively represent steric hindrance and structural intricacy.[57] These customized multidimensional features translate chemical structures of passivators into ML-readable data, laying the foundation for model training.

**2.2. Feature Validation and Machine Learning Modeling**

To assess the efficacy of the proposed 21-dimensional feature set, we employed a Random Forest regression framework as benchmark, and compared its performance against both a reduced 15-feature subset and the descriptor ensembles reported in two representative prior studies (see **Figure S3 and Table S2**). Our full feature set demonstrated superior generalization capability, yielding the highest coefficient of determination on the test set ($R^2$=0.826±0.030) compared to 0.804±0.042 and 0.797±0.044 for Liu et al.'s and Zhi et al.'s feature set. Although the simplified 15-feature subset achieved nearly identical predictive accuracy ($R^2$=0.826±0.031), the complete 21-feature representation was retained to preserve essential physicochemical information required for subsequent interpretability analyses.[58]

Pearson correlation analysis was conducted to evaluate feature redundancy and physical plausibility before ML training (**Figure 2a**). The majority of descriptor pairs exhibited weak correlations ($|r|<0.7$), indicating minimal information overlap. The few instances of strong correlations align well with established physical principles: For instance, the high correlation between the qpmax and EPD ($r=0.87$) reflects the intrinsic coupling between localized charge accumulation and the electrostatic potential distribution, while the moderate correlation between SAmH and SAlH ($r=0.66$) arises from the shared structural motifs among high-performing passivators. Notably, precursor stoichiometric ratios (A:X and B:X) displayed stronger linear associations with I-PCE ($|r| \approx 0.24$) than with the final PCE ($|r| \approx 0.17$), suggesting that effective passivation reduces device performance sensitivity to processing fluctuations. Furthermore, linear correlation analysis underscores the limited predictive power of individual descriptors. While I-PCE and final PCE are strongly positive correlated ($r=0.90$), other molecular features—such as gmean, TPSA, and EPD—exhibit only modest linear trends (Figure 2b). Intriguingly, the molecular dipole moment (u), despite its well-documented role in band alignment in experimental studies, shows negligible linear correlation with device performance (Figure 2a).[59] This apparent inconsistency highlights that passivator efficacy is governed by intricate, nonlinear interactions among multiple physicochemical factors—thereby necessitating ML approaches capable of capturing such higher-order dependencies.[60]

**Table 1**. Description of the 21-dimensional feature set utilized in ML modeling.

|  | Input Features | Feature Description |
| --- | --- | --- |
| *i*) | I_PCE, rA, A:X, B:X | Initial PCE of the control device; Effective radius of the A-site cation; Molar ratios of A-site cation or B-site cation to halide ions in the precursor solution. |
| *ii*) | SHBd, SHBa, SamH, SalH, SarH, SHal | Sum of E-state indices for: hydrogen bond donors, hydrogen bond acceptors, ammonium group hydrogens, aliphatic hydrogens, aromatic hydrogens, and halogens. |
| *iii*) | u, uD, gmean, qpmax, qnmax | Total dipole moment; Dipole moment difference index; Mean square deviation of Gasteiger charges; Maximum positive and negative Gasteiger partial charges. |
| *iv*) | RBR, EPD, LogP, TPSA, vdW, CPLX | Rotatable bond ratio; Maximum electrostatic potential difference; Octanol-water partition coefficient Topological polar surface area; Van der Waals volume; Molecular complexity index. |

To establish a reliable mapping from molecular descriptors to device performance, six ML models, including RF, XGBoost, LightGBM, SVM, KNN, and NN, were systematically evaluated. To improve model robustness against outliers, a hard sample mining strategy was implemented, iteratively reweighting data points with large prediction errors during training (Section S2).[61] As shown in Figure 2c, the RF model consistently outperformed all alternatives, achieving the highest coefficient of determination on the test set ($R^2$=0.914±0.007) alongside the lowest mean absolute error (MAE) and root-mean-square error (RMSE). This represents an ~12.5 % improvement in predictive accuracy compared to the standard training protocol described earlier, corresponding to an absolute $R^2$ gain of 0.1—demonstrating RF's superior capacity to model non-ideal or challenging samples. In contrast, the remaining models exhibited compromised generalization, primarily due to overfitting (e.g., NN, SVM) or underfitting (e.g., KNN), with comprehensive performance metrics provided in the Supporting Information (**Figure S4** and **Table S4**). Consequently, RF was selected as the optimal predictive model for subsequent analyses. Figure 2d further validates the model's fidelity: Predicted PCE values align closely with experimental measurements, with the majority of deviations confined within ±1.0 % (shaded region). Residual analysis confirms that the prediction errors are randomly distributed without evident heteroscedasticity (Figure S4f), satisfying key assumptions of regression reliability. A minor systematic overestimation is observed, likely attributable to subtle discrepancies between idealized computational inputs and real-world experimental conditions; however, this bias remains negligible for high-throughput candidate screening purposes.[62]

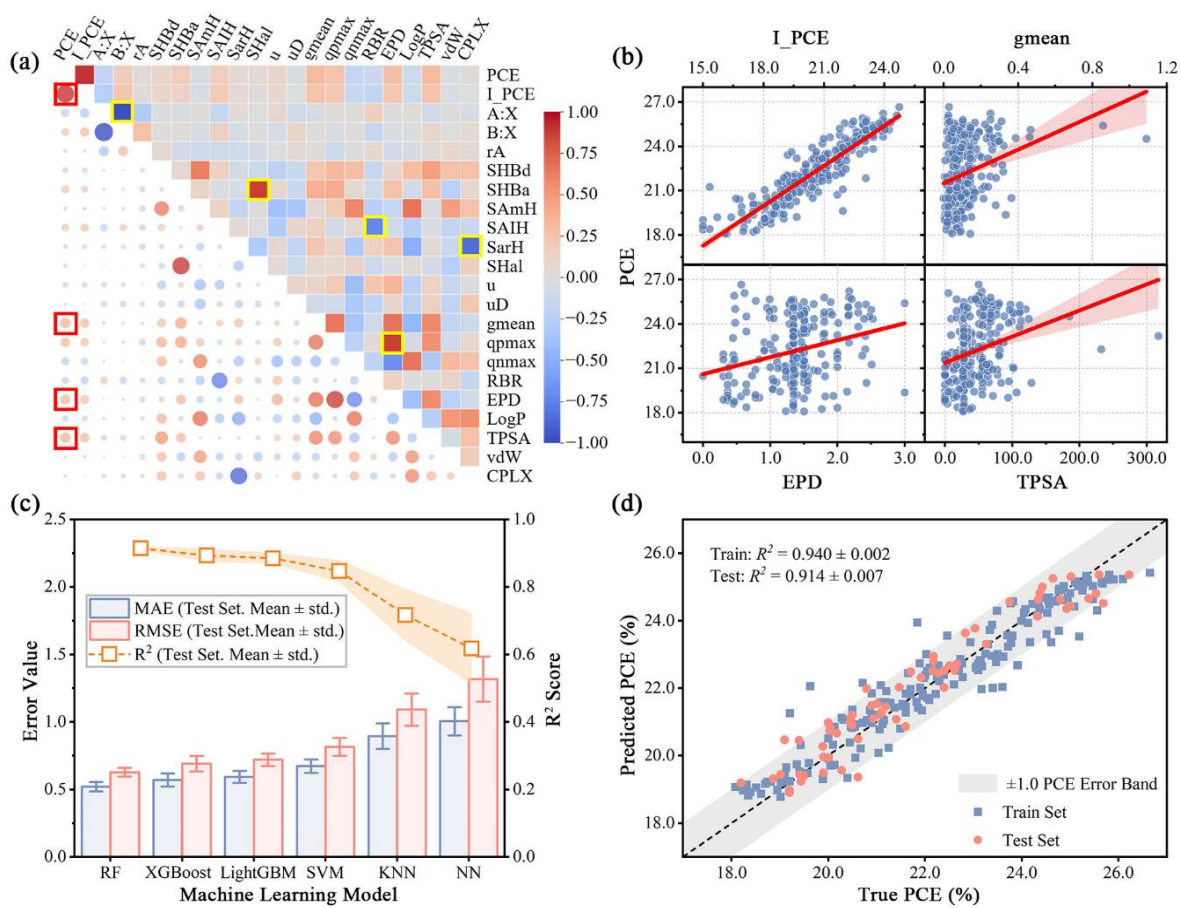

**Figure 2.** Feature correlation analysis and performance evaluation of machine learning models. a) Pearson correlation matrix of the 21 input features. Red boxes highlight descriptors that exhibit strong linear correlation with PCE, while yellow boxes indicate high collinearity (|r|>0.7) among features. b) Linear regression plots showing the relationship between PCE and four key features: I_PCE, gmean, EPD, and TPSA. Shared areas represent 95 % confidence intervals. c) Performance comparison of six regression algorithms on the test set. Error bars denote standard deviations from 10 independent runs. d) Parity plot comparing predicted versus experimental PCE values for the optimized RF model. The gray band represents a ±1.0 % prediction error band.

## 2.3. Interpretability Analysis to Clarify Molecule-Platform Coupling

The built-in importance metric for Random Forest shows the feature ranking, three at top is I_PCE, rA and gmean (**Figure S5**). To uncover the underlying decision logic of the RF model, we performed SHAP analysis.[46] Global SHAP evaluation confirmed that experimental baseline variables—such as I_PCE and rA—exerted dominated influence on model predictions **(Figure S6).** However, to enable rational molecular design, it is essential to isolate the intrinsic contribution of passivators chemistry from these extrinsic experimental factors. Accordingly, we constructed a dedicated SHAP summary plot focusing exclusively on the 17 molecular features, thereby effectively decoupling passivator-specific efficacy from platform-dependent variables—a crucial step for identifying robust chemical drivers independent of fabrication baselines.

As shown in **Figure 3a**, SHBa emerged as the most impactful relationship: Enhanced hydrogen bond accepting capacity consistently correlates with improved PCE. Similarly, polarity-associated features (TPSA, EPD, and u) ranked prominently and exhibited uniformly positive SHAP contributions. Conversely, high LogP values were associated with negative SHAP scores, suggesting that excessive hydrophobicity adversely affects device performance, likely due to limited dispersion in polar precursor solvents.[63] Notably, steric properties also played a non-negligible role: vdW ranked above SHBd in importance and showed a preference for larger molecular volumes. Although SHBd occupies a lower position in the hierarchy, it still displayed a discernible positive trend, suggesting a modest but favorable role in passivation efficacy.

To further investigate how molecular attributes jointly govern performance, we examined non-linear feature interactions using SHAP dependence plots **(Figure S7)**. Four representative pairs were selected to probe cross-domain synergies—gmean vs. vdW, SHBa vs. LogP, RBR vs. u, and uD vs. EPD—charge distribution vs. steric bulk, hydrogen bonding vs. hydrophobicity, conformational flexibility vs. polarity, and electrostatic symmetry vs. strength.

The SHAP-based interaction analysis unveils several actionable molecular design principles. For instance, the SHBa-LogP pair exhibits a "switch-like" dependency: The beneficial effect of strong hydrogen bond acceptor capability is contingent upon a moderately hydrophobic environment—highlighting the context-dependent nature of noncovalent interactions in passivation.[64] Similarly, the interplay between gmean and vdW interaction suggests that optimal performance arises from a delicate balance between electronic heterogeneity and moderate steric bulk, avoiding extremes that could disrupt film morphology or charge transport.[65,66] Additionally, molecular flexibility (RBR) was found to attenuate the typically detrimental impact of large dipole moments (u), likely by facilitating conformational adaptation at interfaces to minimize energetic penalties.[67] A comprehensive discussion of these cooperative mechanisms is provided in Note S5 and visualized in Figure S7.

Beyond intrinsic molecular features, the model also uncovered delicate dependencies between passivation efficacy and key experimental variables (**Figure S8**). The influence of parameters such as rA, A:X, and B:X on final PCE is distinctly non-monotonic and strongly modulated by initial device quality (I_PCE). Specifically, the positive contribution of rA is observed only within a narrow window (2.4–2.6 Å), aligning closely with the ideal range predicted by the Goldschmidt tolerance factor for stable perovskite lattice.[68] Furthermore, deviations from nominal stoichiometric—particularly Pb-rich conditions—trend to enhance performance in initially high-quality films but degrade those with poor baseline characteristics. This dichotomy provides computational support for targeted "Pb-rich" processing strategies, which appear most effective when applied to already optimized device architectures.[69-71]

To bridge global feature-level insights with individual predictions, we generated a SHAP heatmap (Figure 3b) that visualizes the contribution of each descriptor across all 240 samples, ordered by predicted PCE. The heatmap demonstrates that high-performance predictions arise from the concerted positive contributions of multiple features (evident as deep red blocks), reinforcing the notion that effective passivation is governed by synergistic, multivariate effects rather than isolated descriptors.

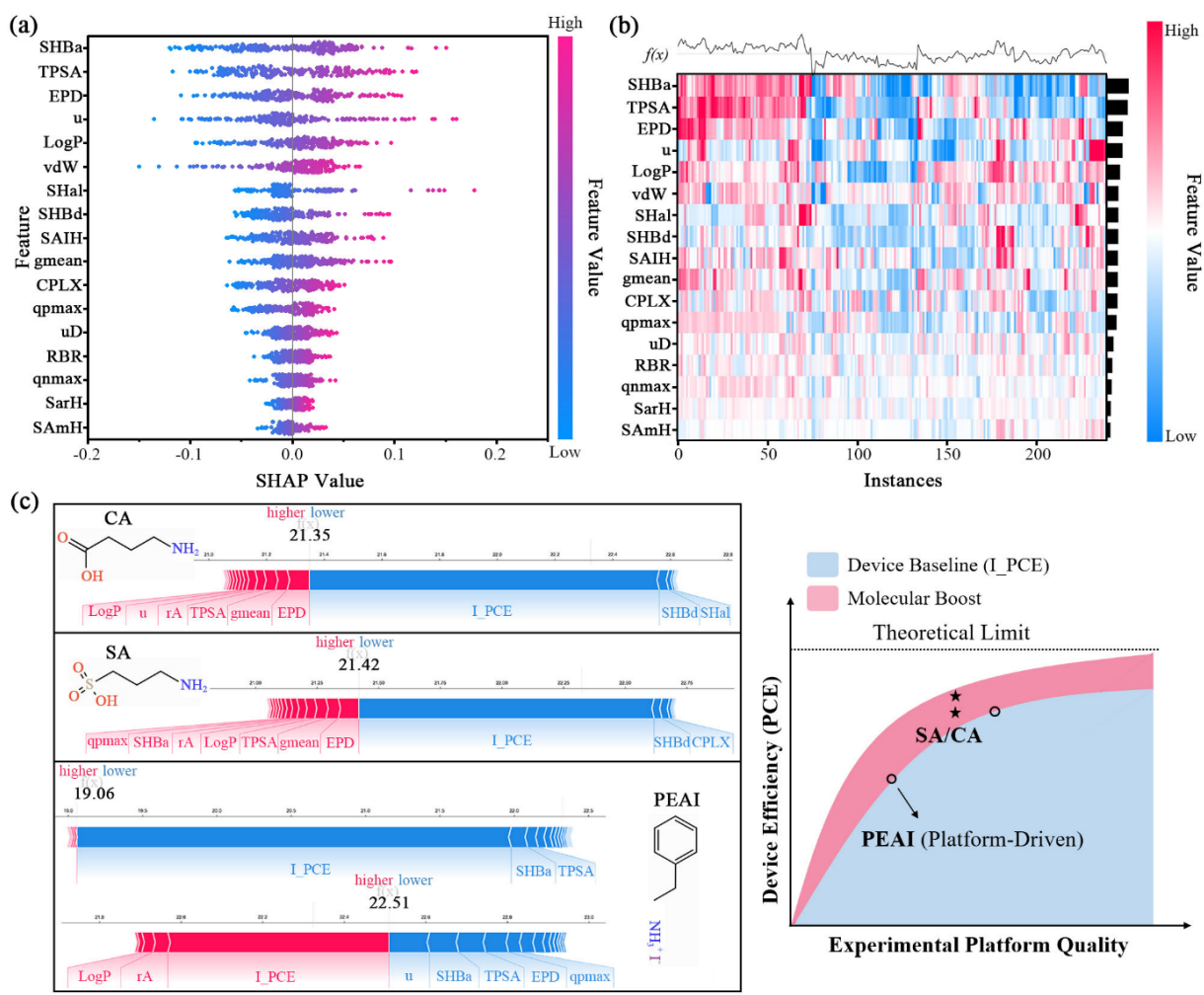

**Figure 3.** Multi-level interpretability analysis using SHAP. a) Summary plot illustrating the global importance and directional impact of the 17 molecular features. b) Heatmap of feature contributions across all 240 samples, sorted by model output f(x). c) Local attribution and mechanistic decoupling. Left: SHAP force plots comparing CA vs. SA to elucidate the molecular effect, and PEAI under different baselines conditions to illustrate platform effect. Right: Schematic representation of an asymptotic saturation model distinguishing the "Intrinsic Molecular Boost" (pink zone, stars) from "Platform-Driven" efficacy (blue baseline region, circles) relative to the theoretical limit.

To further dissect these contributions at the molecular level, SHAP force plots were employed for analyzing representative passivators (Figure 3c, left). A direct comparison between CA and SA reveals that the sulfonyl group ($-SO_2-$) in SA substantially enhances the SHAP values of SHBa and qpmax relative to the carboxyl group in CA—thereby accounting for its higher predicted efficiency.[72-74] In contrast, PEAI exhibits consistently negative or negligible contributions from its molecular features (e.g., u, SHBa), irrespective of device quality. This indicates that its observed PCE improvement stems predominantly from a favorable experimental baseline (I_PCE), underscoring the strong interdependence between passivator chemistry and the underlying device platform.[75-77] Collectively, these cases studies confirm that model predictions are chemically interpretable and aligned with established structural-function relationships.

To quantitatively disentangle the roles of molecular design and experimental context, we introduced an asymptotic saturation model (Figure 3c, right). In this framework, the total PCE enhancement follows a non-linear saturation curve approaching a theoretical upper limit. The blue region represents the performance floor established by the experimental platform (I_PCE), while the superimposed pink zone—termed "Molecular Boost"—captures the additional gain attributable to intrinsic molecular properties. Notably, SA and CA (marked by stars) lie well within the pink zone, confirming that their functional groups (e.g., -$SO_2$-) confer genuine chemical advantages beyond platform effects. Conversely, PEAI (circles) closely adheres to the baseline boundary, consistent with its weak SHAP feature scores and implying that its efficacy is largely platform-dependent. High final PCE derives primarily from the high-quality platform rather than intrinsic chemical. This decomposition highlights a critical design principle: Future high-impact passivators must be engineered to actively penetrate the "Molecular Boost" regime, rather than relying on fortuitous compatibility with high-performing platforms. By explicitly separating these contributions, our framework ensures the unbiased discovery of molecules with true intrinsic gains, overcoming the confounding factors that often obscure chemical advances in literature.

## 2.4. Large-Scale Virtual Screening and Candidate Discovery

leveraging the validated predictive model and the mechanistic insights derived from SHAP analysis, we implemented a scalable framework for high-throughput molecular discovery. A hierarchical virtual screening pipeline was developed to interrogate the PubChem database—encompassing over 121 million compounds—for promising passivator candidates (**Figure 4a**). The workflow comprised three sequential filtering steps. Step 1 employed computationally lightweight, formula-based criteria—including MW, DoU, elemental ratios—to exclude thermally unstable, synthetically intractable, or overly complex structures. This initial curation reduced the candidate pool to approximately 102 million molecules. In Step 2, we introduced a chemically informed "dual-functional" motif filter based on SMILES pattern matching. Candidates were required to simultaneously harbor both electron-donating groups (e.g., H-bond donor) and electron-accepting moieties (e.g., Lewis base). This design principle targets the coexistence of Lewis acidic sites (e.g., undercoordinated $Pb^{2+}$) and Lewis basic defects (e.g., halide vacancies) commonly found on perovskite surfaces, thereby enabling multidentate, synergistic passivation through cooperative binding.[78-80] Application of this constraint narrowed the library to 1.38 million structurally plausible candidates. Step 3 incorporated feature-based physicochemical optimization. Key 2D molecular features—related to polarity (TPSA), LogP, steric size, conformational flexibility, and topological complexity—were computed and subjected to empirically grounded bounds. These constraints, while avoiding the computational overhead of the full 21-dimensional feature set, were directly informed by trends identified in our SHAP interpretability analysis. Analogous to "drug-likeness" rules in medicinal chemistry, these filters ensure favorable solution processability, interfacial compatibility, and film-forming behavior.[81] This three-tiered filtration strategy ultimately yielded a refined library of 789,931 high-potential passivators suitable for downstream predictive ranking. Full screening criteria and parameter thresholds are detailed in Note S6.

Using the trained RF model, we evaluated the performance of the curated library under three virtual experimental scenarios. All scenarios identical system-level parameters—namely rA=2.48, A:X=0.34, B:X=0.33—extracted from high-performance samples in our dataset, while varying the initial baseline quality (I_PCE) to reflect the 60th (21.09 %), 80th (22.60 %), and 95th (23.51 %) percentiles of the baseline distribution. As shown in Figure 4b, the predicted PCE distribution exhibited clear sensitivity to I-PCE: Higher-quality starting platforms not only shifted the entire distribution upward but also reduced its variance, indicating more consistent and reliable passivation outcomes on optimized substrates.

To prioritize candidates for downstream validation, we devised a selection strategy that jointly optimizes predictive performance and structural diversity. From each scenario, the top 5,000 high-scoring molecules were extracted. Molecules appearing in all three scenario-specific top lists were designated as core candidates, reflecting robustness across varying baseline conditions. The remaining 12,888 unique high-performers—present in one or two scenarios—were subjected to diversity-aware filtering to avoid chemical redundancy. Structurally similarity was quantified using Extended Connectivity Fingerprints (ECFPs), and the Butina clustering algorithm was applied to partition the set into structurally coherent groups.[82-84] t-SNE embedding of the fingerprint space (Figure 4c) reveals that high-performing candidates populate multiple well-separated clusters, underscoring significant scaffolds diversity. The five largest clusters were identified, and their representative molecular scaffolds are highlighted for clarity.[85,86] From each cluster, the top 5 molecules were selected based on a weighted ranking score that integrates predicted PCE, SHAP-based feature favorability, and synthetic accessibility (see Note S8). This approach ensures broad coverage of chemical space while mitigating overrepresentation of any single structural motif.

The 1,201 structurally diverse representative molecules were prioritized using the "Expected Improvement Ratio" (EIR), defined as the predicted PCE devided by I_PCE. As shown in Figure 4d, their distribution across medium- and high-performance scenarios reveals that the newly screened candidates (green dots) consistently occupy regions of higher potential compared to re-evaluated known passivators (orange crosses), thereby validating the efficacy of the discovery workflow. Nevertheless, reliance on point predictions alone entails inherent risk due to model uncertainty.[87,88] Without uncertainty filtering, ~40 % of the top-100 candidates exhibit wide prediction intervals (> ±0.05 EIR width), indicating a high risk of overestimation and potential false positives. To address this, a Random Forest Quantile Regression model was trained to estimate the 90 % prediction interval for the EIR (see Note S7). Candidates were then filtered using a dual criterion: High expected improvement coupled with low predictive uncertainty (i.e., narrow confidence intervals). This strategy yielded five top-performing molecules, highlighting as red circles in Figure 4d.

Figure 4e presents a performance comparison between the five-top ranked candidates and established experimental benchmarks—PEAI and tBBAI. Across all virtual scenarios, the selected candidates consistently deliver improvement ratio exceeding 1.08, surpassing both reference passivators in predicted efficacy. For clarity and conciseness, each candidate is labeled according to its core heterocyclic scaffold: TDZ-S (thiadiazolidine dioxide), TZC-F and TZC-P (thiazolidine derivatives functionalized with difluoro and pivaloyl groups, respectively), DZP-A (diazepane-based architecture), and ODZ-F (fluorinated oxadiazole). Complete IUPAC names, chemical structures, and associated descriptors are provided in **Table S5**. Importantly, automated retrosynthesis analysis via ChemAIRS confirms that all five finalists are either from commercially available precursors or accessible in 2-4 synthetic steps (see Note S8 and **Figure S10**), underscoring their experimental feasibility. As the principal output of our data-driven discovery pipeline, these molecules represent high-value, chemical diverse leads for synthesis and device-level validation.

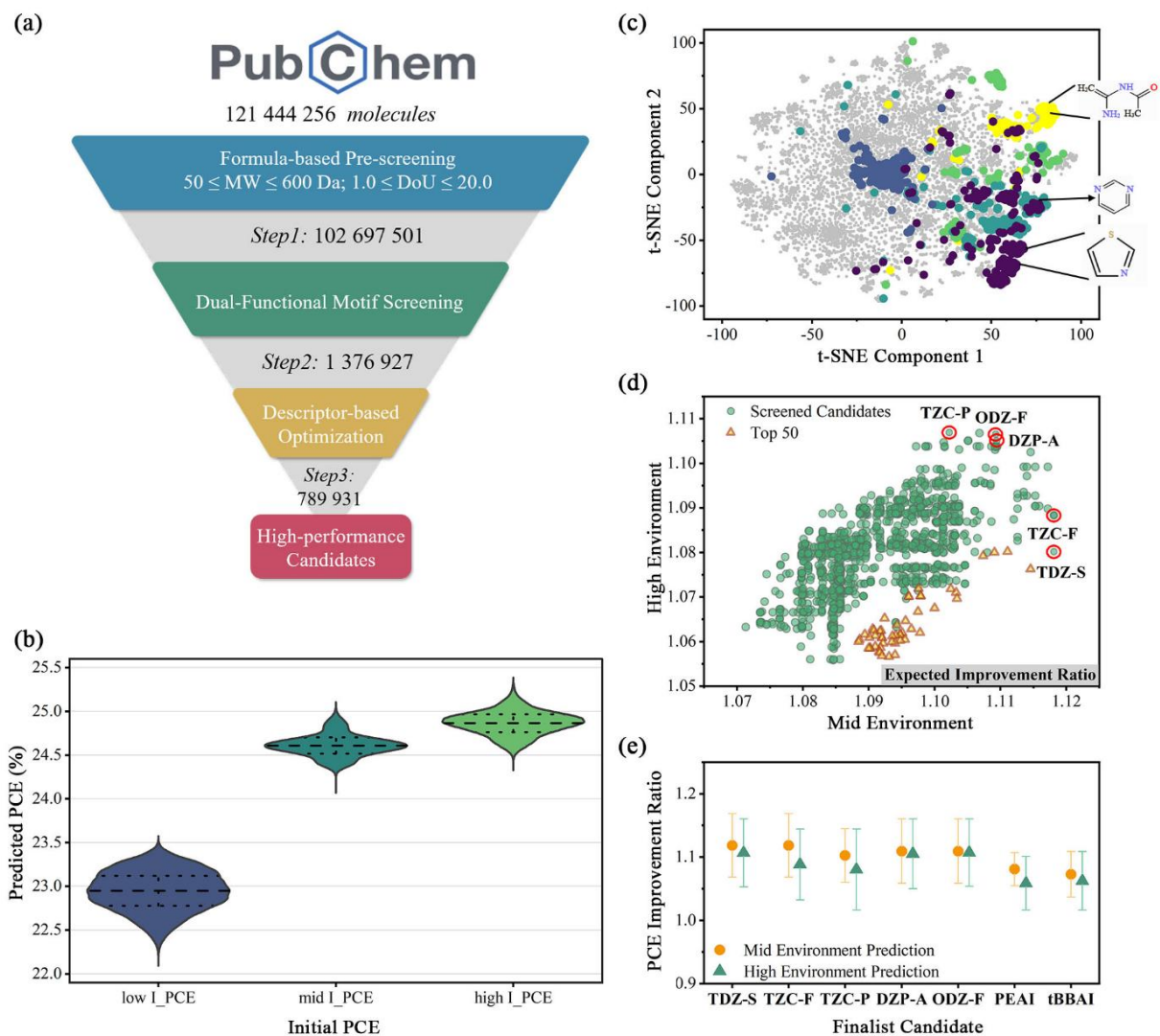

**Figure 4.** High-throughput virtual screening and discovery of candidate passivators. a) Schematic of the hierarchical screening workflow. The pipeline reduces the chemical space from over 121 million PubChem entries to a refined library of high-pitential candidates through a series of filters. b) Violin plots of predicted PCE distributions. These plots show the predicted PCE distributions for the refined library across three virtual scenarios. c) t-SNE visualization of the chemical space. The plot visualizes the chemical diversity of elite candidates based on ECFP fingerprints. d) Expected PCE improvement ratios. The plot compares the expected improvement ratios for screened candidates (green dots) versus training set molecules (orange crosses), with top five finalists highlighted in red circles. e) Performance comparison and uncertainty quantification. This panel evaluates the performance and uncertainty of the five finalists against experimental benchmarks (PEAI, tBBAI) under medium and high-performance conditions.

## 2.5. First-Principles Verification of Atomistic Mechanisms

To uncover the atomistic and electronic underpinnings of the predicted performance, we carried out first-principles calculations (see Note S9). We began by analyzing the electrostatic potential (ESP) mapped onto the electron density isosurface, as shown in **Figure 5a**. A consistent electronic motif emerged across all five top candidates: Pronounced charge inhomogeneity featuring spatially separated electron-rich (red) and electron-deficient (blue) regions. This intrinsic polarization aligns directly with SHAP-based insights that identified high gmean and strong EPD as critical performance drivers.

Notably, strong negative ESP minima were localized on electronegative functional groups—such as sulfonyl (-$SO_2$-) and carbonyl (-C=O) oxygen atoms in TDZ-S and TZC-F. These sites serve as potent Lewis bases and hydrogen bond acceptors, corroborating the high SHAP importance assigned to SHBa. Analogous electron-rich centers were observed on amide oxygens and heterocyclic atoms in DZP-A (a piperazine derivative) and ODZ-F (an oxadiazole derivative), reinforcing their capacity for anion coordination. Conversely, pronounced positive ESP maxima were found on amine or amide protons (-NH), particularly in TDZ-S and DZP-A. These electropositive sites function as hydrogen bond donors, enabling favorable interactions with surface halide vacancies—a role consistent with the positive (though secondary) contribution of SHBd in our interpretability analysis. These electronic features establish a "dual-functional" molecule architecture: Each candidate simultaneously presents complementary Lewis acidic and basic sites, enabling cooperative, multi-point passivation of coexisting $Pb^{2+}$ halide defects. This atomistic picture provides a mechanistic foundation for the superior performance predicted by our ML model, confirming that the high predicted efficacy stems from intrinsic molecular capabilities rather than platform dependencies.

The binding affinity of the top candidates to the β-$CsPbI_3$(001) surface—terminated with $PbI_2$ layer—was quantified via adsorption energies ($E_{ads}$). Multiple adsorption configurations were evaluated, with particular focus on energetically favorable "parallel" and "vertical" orientations (Figure 5b). All five molecules exhibited strong chemisorption, with $E_{ads}$ values below −1.7 eV—significantly more exothermic than typical physisorption thresholds (< 0.5 eV). Notably, TZC-F and DZP-A achieved exceptionally stable binding, with $E_{ads}$ being −2.73 eV and −2.63 eV, respectively, underscoring their robust interfacial anchoring capability.

This chemisorption profoundly modulates the surface electronic structure. As shown in the energy level alignment diagram (Figure 5c), molecular adsorption induces a pronounced shift in the Fermi level ($E_F$) and a marked reduction in the work function (Φ). For instance, TZC-F lowers Φ from 5.38 eV (pristine surface) to 5.15 eV. Such a reduction improves band alignment with common electron transport layer, thereby diminishing the interfacial barrier and promoting efficient electron extraction.[48] Concurrently, the positions of the valence band maximum (VBM) and conduction band minimum (CBM) are subtly shifted upon adsorption, potentially enhancing interfacial carrier dynamics and reducing recombination losses.[89,90]

To elucidate the bonding mechanism, we performed differential charge density analysis (Figure 5d). The 3D isosurfaces (upper panels) visualize interfacial charge redistribution: Yellow regions denote electron accumulation, while cyan regions indicate depletion. Clear charge transfer channels emerge across all systems. Taking TZC-F as an example, significant electron accumulation occurs between the sulfonyl oxygen and undercoordinated surface Pb atoms, accompanied by electron depletion around Pb sites—characteristic of coordinative bonding. This confirms that the ESP-identified electron-rich centers serve as active anchoring sites for Lewis acidic $Pb^{2+}$. Analogous charge transfer patterns are consistently observed for the sulfonyl group in TDZ-S, the carbonyl (-C=O) moiety in TZC-P, and the amide oxygen in DZP-A, collectively validating the dual-functional design principle at the atomic scale.

Quantitative analysis of the planar-averaged differential charge density ($\Delta\rho$, green curves) revealed pronounced interfacial oscillations (0.1−0.2 $e/Å^3$), indicative of strong chemical bonding at the molecule-perovskite interface. Complementarily, integrated charge displacement profiles ($\Delta Q$, red curves) reveal that all five candidates exhibit negative plateau values in $\Delta Q$ ( −0.05 to −0.15 $e$), conforming their role as net electron donors to the perovskite lattice. This unidirectional charge transfer directly passivates electron-deficient surface defects—particularly undercoordinated $Pb^{2+}$ ions—thereby neutralizing deep-level trap states that drive non-radiative recombination centers.[14,91] Critically, this consistent electron-donating behavior emerged organically from our data-driven discovery pipeline: The machine learning framework converged on molecules with this specific electronic character without explicit quantum-chemical constraints, demonstrating its capacity to uncover latent structure-function relationships and identify intrinsic passivation mechanisms. The resulting Lewis acid-base interaction—where electron-rich functional groups on the passivators donate charge to Lewis acidic $Pb^{2+}$ sites—forms robust coordinative bonds that accounts for both the high adsorption energies and the favorable interfacial energy alignment. These atomistic insights provide a mechanism foundation fort the experimental prioritization of these candidates for synthesis and device integration.

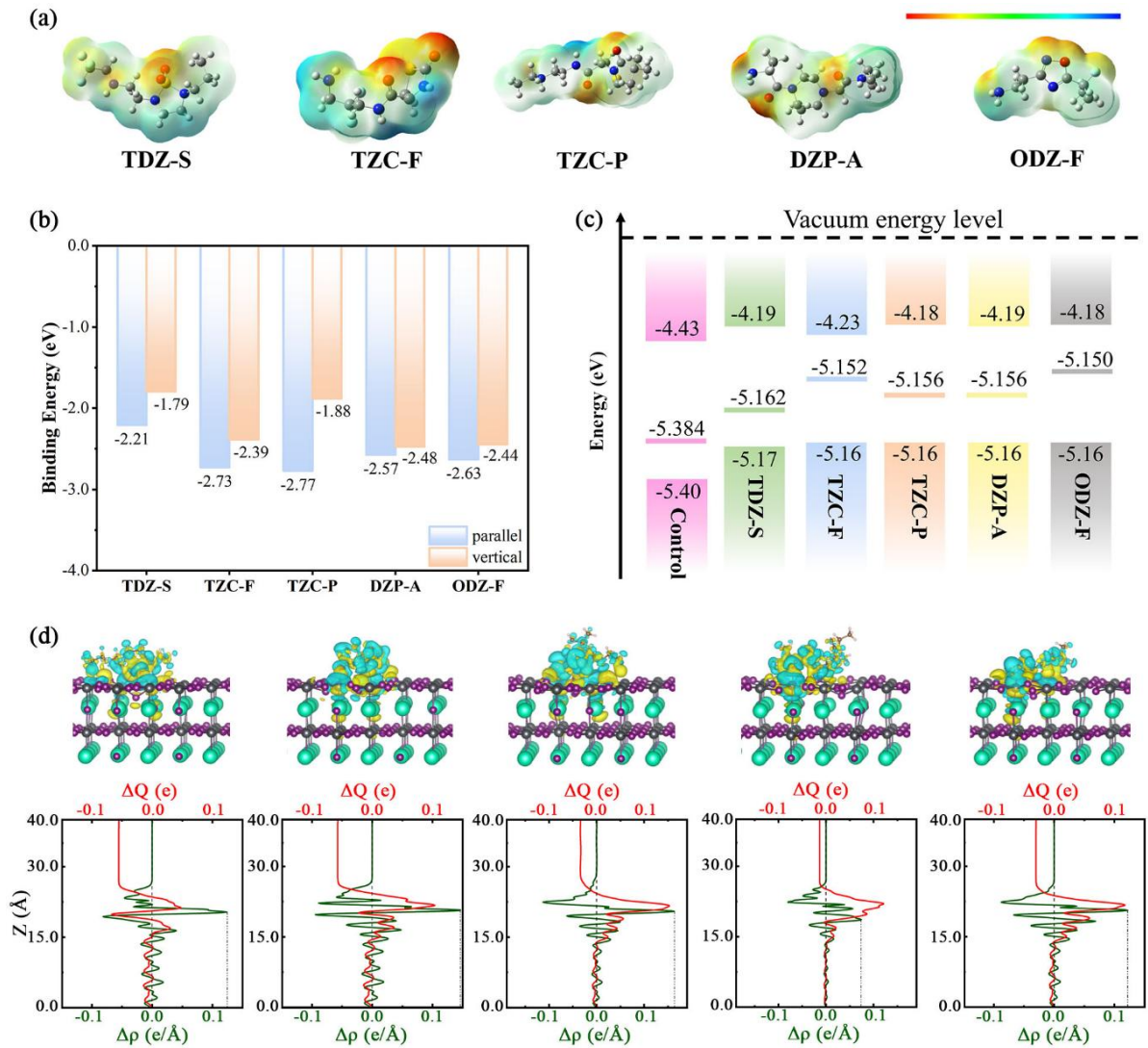

**Figure 5.** First-principles validation of the finalist candidates. a) Electrostatic potential (ESP) maps mapped onto the electron density isosurface (0.002 a.u.). b) Calculated $E_{ads}$ for the five candidates on the β-CsPbI$_3$(001) surface, comparing parallel and vertical configurations. c) Energy level alignment diagram showing the shift in Fermi level ($E_F$) and work function (Φ) upon molecule adsorption relative to the pristine control. d) Differential charge density analysis for the most stable adsorption configurations. Upper panels: 3D isosurfaces (yellow: accumulation; cyan: depletion). Lower panels: Planar-averaged charge density difference (Δρ, green curves) and integrated charge displacement (ΔQ, red curves) along the z-axis.

## 3. Conclusion

In conclusion, we establish a generalizable, interpretable machine learning framework that decouple intrinsic molecular efficacy from platform-dependent performance—a long standard challenge in rational passivator design for perovskite interface. Built on 240 experimental validated entries and a physically informed 21-dimensional feature set, our hard sample mining strategy Random Forest model achieves high predictive accuracy ($R^2$=0.914). SHAP-based interpretability analysis reveals hydrogen bond acceptor strength (SHBa and electrostatic potential difference (EPD) as key molecular determinants for intrinsic passivation efficacy. Crucially, we introduced an asymptotic saturation model that quantitatively separates "Platform-Driven" baseline performance from the "Intrinsic Molecular Boost". This analysis demonstrates that conventional agents like PEAI largely mirror substrate quality, whereas sulfonic (SA) and cinnamic acid (CA) deliver genuine chemical enhancements that push device efficiency toward its theoretical limit—the "pink zone". Together with a newly identified "hydrophobicity-gated" interaction mechanism, these insights yield actionable, substrate-independent design rules.

Guided by these principles, we perform hierarchical virtual screening across >121 million molecules in PubChem, integrating dual-functional motif constraints, SHAP-informed filters, diversity clustering, and uncertainty-aware ranking. This yields five high-priority candidates—TDZ-S, TZC-F, TZC-P, DZP-A, and ODZ-F—All predicted to enhance device performance ratio greater than 1.08 relative to state-of-art benchmarks. First-principles calculations validate their "dual-functional" architectures: Spatially separated Lewis basic/acidic sites enable strong chemisorption ($E_{ads}$ < −1.7 eV), net electron donation (–0.05 to –0.15 $e$), and effective passivation of undercoordinated $Pb^{2+}$ defects, while concurrently optimizing interfacial energy alignment for efficient charge extraction.

Beyond delivering synthetically accessible leads, this work establishes a close-loop "data–interpretation–screening–verification" paradigm for materials discovery. It demonstrates how physically grounded, interpretable ML can unravel complex structure-property relationships and accelerate rational design—an approach readily extensible to diverse interface engineering challenges, including hole/electron transport layer optimization and 2D perovskite modulation, as well as other optoelectronic devices such as OLEDs, quantum dot devices.

**Acknowledgements**

This work is supported by the National Natural Science Foundation of China (Grant No. 12374061, No. 12404274, No. 12204256), Zhejiang Provincial Natural Science Foundation of China (Grant No. LQN25A040020, No. LD24F040001), the KC Wong Magna Foundation in Ningbo University. The computations were supported by high performance computing center at Ningbo University.

**Conflict of Interest**

The authors declare no competing financial interest.

**Data Availability Statement**

The data that support the findings of this study are available from the corresponding author upon reasonable request.

Received: ((will be filled in by the editorial staff))
Revised: ((will be filled in by the editorial staff))
Published online: ((will be filled in by the editorial staff))

Semiconductor Processing 204 (2026): 110278.

https://doi.org/10.1016/j.mssp.2025.110278

[65] S. Teale, M. Degani, B. Chen, et al., "Molecular Cation and Low-Dimensional Perovskite Surface Passivation in Perovskite Solar Cells," *Nature Energy* 9 (2024): 779–792.

https://doi.org/10.1038/s41560-024-01529-3

[66] Y. Miao, Y. Chen, H. Chen, X. Wang, and Y. Zhao, "Using Steric Hindrance to Manipulate and Stabilize Metal Halide Perovskites for Optoelectronics," *Chemical Science* 12 (2021): 7231–7247.

https://doi.org/10.1039/D1SC01171E

[67] Y.-W. Yang, K. Xu, Z.-N. Zhou, M.-L. Jin, R. Tsunashima, T. Nakamura, C.-Y. Chai, and Q. Ye, "Organic Cation Conformational Flexibility Governs Mechanical Response in Organic-Inorganic Hybrid Materials," *Chemical Science* (2026): Advance Article.

https://doi.org/10.1039/D5SC06333G

[68] C. J. Bartel, et al., "New Tolerance Factor to Predict the Stability of Perovskite Oxides and Halides," *Science Advances* 5 (2019): eaav0693.

https://doi.org/10.1126/sciadv.aav0693

[69] L. Zhang, W. Wang, Y. Wei, H. Wang, J. Ye, P. Lin, P. Wang, X. Wu, X. Yu, Z. Ni, L. Xu, and C. Cui, "Assessing the Effect of Excess $PbI_2$ on the Photovoltaic Performance of $CsPbI_3$ All-Inorganic Perovskite Solar Cells," *Materials Today Communications* 46 (2025): 112548.

https://doi.org/10.1016/j.mtcomm.2025.112548

[70] S. Li and J. T. S. Irvine, "Non-Stoichiometry, Structure and Properties of Proton-Conducting Perovskite Oxides," *Solid State Ionics* 361 (2021): 115571.

https://doi.org/10.1016/j.ssi.2021.115571

[71] X. Qi, T. Zhang, F. Tan, et al., "Self-Passivated Hybrid Perovskite Films for Improved Photovoltaic Performance of Solar Cells," *Journal of Materials Science* 56 (2021): 6374–6384.

https://doi.org/10.1007/s10853-020-05685-1

[72] J. Fang, L. Wang, Z. Chen, S. Wang, L. Yuan, A. Saeed, I. Hussain, J. Zhao, R. Liu, and Q. Miao, "Sulfonic Acid Functionalized Ionic Liquids for Defect Passivation via Molecular Interactions for High-Quality Perovskite Films and Stable Solar Cells," *ACS Applied Materials & Interfaces* 16 (2024): 23443–23451.

https://doi.org/10.1021/acsami.4c04762
28

**Supporting Information**

Supporting Information is available from the Wiley Online Library or from the author.

**Decoupling Intrinsic Molecular Efficacy from Platform Effects: An Interpretable Machine Learning Framework for Unbiased Perovskite Passivator Discovery**

This study establishes an interpretable machine learning framework that disentangles the intrinsic molecular efficacy of passivators from experimental platform effects—enabling unbiased, high-throughput discovery of effective perovskite surface modifiers. By integrating a data-driven asymptotic saturation model with uncertainty-quantified virtual screening across 121 million molecules, we identify five dual-functional candidates with strong predicted performance. This closed-loop strategy transforms materials design from empirical trial-and-error to a rational, mechanism-guided paradigm, accelerating the discovery of next-generation perovskite solar cell passivator.

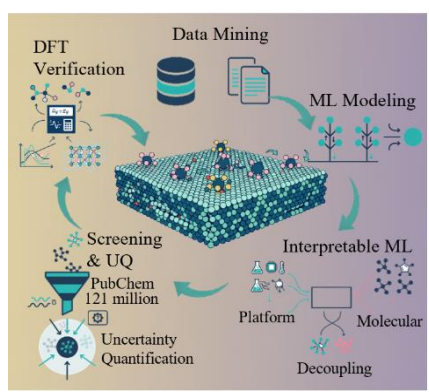